\documentclass[reprint, superscriptaddress, aps,longbibliography]{revtex4-2}
\usepackage{graphicx}    
\usepackage{dcolumn}     
\usepackage{bm}          
\usepackage{color}
\usepackage{amsmath, amssymb}
\usepackage{newtxtext, newtxmath}
\usepackage{mathrsfs}
\usepackage{comment}

\newcommand{\braket}[1]{\left\langle #1 \right\rangle}
\newcommand{\Bra}[1]{\left\langle #1 \right|}
\newcommand{\Ket}[1]{\left| #1 \right\rangle}

\begin{document}
\title{First-principles method justifying the Dieke diagram and beyond}
\author{Katsuhiro~Suzuki} 
\affiliation{Division of Materials and Manufacturing Science, Graduate School of Engineering, Osaka University, Suita, Osaka 565-0871, Japan}
\author{Takao~Kotani}
\affiliation{Advanced Mechanical and Electronic System Research Center (AMES), Faculty of Engineering, Tottori University, Tottori 680-0945, Japan}
\affiliation{CSRN-Osaka, Osaka University, Toyonaka, Osaka 560-8531, Japan}
\author{Kazunori~Sato}
\affiliation{Division of Materials and Manufacturing Science, Graduate School of Engineering, Osaka University, Suita, Osaka 565-0871, Japan}
\affiliation{CSRN-Osaka, Osaka University, Toyonaka, Osaka 560-8531, Japan}
\affiliation{Spintronics Research Network Division, OTRI, Osaka University, Toyonaka, Osaka 560-8531, Japan}
\date{\today}
\begin{abstract} 
    We present a method to determine the model Hamiltonians to treat rare-earth multiplets in solids from the results of the quasiparticle self-consistent \textit{GW} (QSGW) method. We apply the method to trivalent Eu compounds EuCl$_3$, EuN, and Eu-doped GaN after examining free rare-earth ions. We solve the model Hamiltonian by the exact diagonalization. Our results justify applying the Dieke diagram to ions in solid, while its limitation is clarified. In particular, we show that the crystal fields cause sizable breaking of the Russell-Saunders coupling.
\end{abstract}
\maketitle
\section{Introduction}
The photoluminescence of rare-earth atoms (PLR) embedded in solids is one of the essential physical phenomena, which has a wide range of possible applications for light emitters. For example, Eu-doped GaN is a candidate for trichromatic LEDs, thanks to red-light emission by PLR of  Eu\cite{1039467, doi:10.1063/1.3478011, doi:10.1063/1.1821630}. 
It is favorable to assist the investigation of such PLR with the computations,
that is, the computational materials design (CMD). However, CMD for PLR is not so easy because we do not have a reliable theoretical basis to perform computations for PLR even today. 

Historically, the model-Hamiltonian methods based on the atomic multiplet theory had been developed to explain complex spectra of PLR, where the model Hamiltonian was parametrized by the Slater integrals \cite{PhysRev.34.1293},
which are treated as material-dependent parameters.
The atomic multiplet theory to determine the model Hamiltonian was given by Racah\cite{PhysRev.76.1352}, followed by further developments \cite{doi:10.1063/1.1732879,doi:10.1063/1.1733947,doi:10.1063/1.1669893,PhysRev.127.750,doi:10.1063/1.1701366}. 
Particularly, a comprehensive study by Dieke and Crosswhite on doubly and triply ionized rare earth (\textit{RE}) is well known. The energy level diagram for all lanthanides in the study called the Dieke diagram is taken as a standard for analyzing experimental results even now\cite{Dieke:63}. The material-dependent parameters 
to give the Dieke diagram are determined experimentally so as to reproduce the optical spectra of PLR. Since we had no means to determine the parameters by computation, the model-Hamiltonian methods were quite limited from the view of CMD. 

One of the recent theoretical approaches to treating the multiplets was given by Ungur and Chibotaru, who successfully applied the complete active space self-consistent-field method (CASSCF) \cite{doi:10.1063/1.441359} to lanthanide complexes \cite{chem.201605102,acs.inorgchem.2c00071}.
However, applying CASSCF to handle \textit{RE} in solids such as semiconductors should be difficult. This is because we have to consider the screening effects of the Coulomb interaction between $4f$ electrons caused by the polarization of host semiconductors. Furthermore, $4f$ orbitals can hybridize well with atomic orbitals in solids: the $4f$ orbitals can be itinerant or localized depending on their environments \cite{PhysRevLett.43.1892,PhysRevLett.127.067002,CeRh2Si2ARPES,PhysRevLett.105.237601}. 
We have to use a method reproducing the properties of both \textit{RE}
and solids on the same footing. In CASSCF, it is hopeless from the view of high computational demand to use sufficiently large enough active space to reproduce the properties of solids. CASSCF is applicable to only the cases where $4f$ electrons are very localized.

We have other first-principles-based methods to handle the multiplets until now. Most of these methods have two key steps. One is how to determine the model Hamiltonian, and the other is how to solve the model, namely the solver. For the solver, we have extensive developments until now. Assuming the LDA+$U$ model Hamiltonian, the Hubbard-I approximation\cite{doi:10.1098/rspa.1963.0204, PhysRevB.57.6884} (HIA) and the dynamical mean-field theory (DMFT) have been developed to predict multiplet properties and peak structures of the excitation spectra\cite{PhysRevB.71.045119, PhysRevB.72.245102, PhysRevB.74.045114, PhysRevB.100.161104}.

A serious problem is in the first step. The reliability of widely used LDA+$U$ is quite questionable. We usually have to give the size of $U$ by hand. LDA+$U$ assumes a very simple double-counting term. There are no parameters to control the center of $4f$ bands relative to the anion bands, that is, we simply assume that the center is determined in LDA. Considering these facts, we guess that LDA+$U$ is often abused with little theoretical justifications as was analyzed by Lee, Kotani, and Ke\cite{lee_role_2020}. Moreover, the multiplet excitation of \textit{RE} is not controlled by $U$ but by the 2nd Slater integral $F_2$ as we will explain later on. 

In this paper, we will give a new method for the first step, deducing the parameters in the model Hamiltonian from the results of the quasiparticle self-consistent \textit{GW} (QSGW) method \cite{PhysRevB.76.165106}. QSGW is a reliable mean-field approximation in the sense that the one-particle Hamiltonian determined in QSGW gives a good independent-particle picture for a wide range of materials\cite{1347-4065-55-5-051201,PhysRevB.76.165106} including not only semiconductors but also $4f$ systems \cite{PhysRevB.76.165126}.
QSGW is roughly identified to be a ``screened'' Hartree-Fock method where	the screening effect is internally determined self-consistently.
Both excitation energies and quantities such as spin fluctuations can be reproduced well based on the one-particle Hamiltonian in QSGW \cite{doi:10.7566/JPSJ.90.094710}. 
Let us explain our core idea about how to deduce the parameters in the model Hamiltonian. Our core idea is that we require ``QSGW applied to the model Hamiltonian'' should reproduce the (model part of) one-body Hamiltonian given in QSGW. In contrast to LDA, we have no theoretical problem applying QSGW to the model Hamiltonian.

In this paper, we apply our new method to the multiplets of $4f$ orbitals in \textit{RE} compounds, EuCl$_3$, EuN, and Eu-doped GaN, after examining trivalent \textit{RE} ions (\textit{RE}$^{3+}$) in the supercell. The parameters of the model Hamiltonian for describing $4f$ electrons are obtained based on our core idea. With the
exact diagonalization applied to the model Hamiltonian, we show the eigenvalues of multiplets of \textit{RE} $4f$ orbitals and discuss the relation with experiments.

\section{Method}
We assume a model Hamiltonian of the multiplets with a fixed number of electrons of $4f$ orbitals. Here we neglect the hybridization of $4f$ orbitals with the other orbitals. Thus, the dimension of the model Hamiltonian is the number of the electronic configurations of $4f$ electrons, i.e., $_{14}C_n$ where $n$ is the number of $4f$ electrons. Then, the model Hamiltonian is written as
\begin{align}
  \mathscr{H}&=\mathscr{H}_\textrm{0}+\mathscr{H}_\textrm{CF}+\mathscr{H}_\textrm{SOC}+\mathscr{H}_\textrm{C}. \label{eq:hamiltonian}
\end{align}
$\mathscr{H}_\textrm{0}$ is a constant matrix to give the base level of $4f$. The level is irrelevant to energy spectra since we do not consider the hybridizations. The non-spherical part of the one-body potential is given by the crystal field term $\mathscr{H}_\textrm{CF}$. In addition, we have the spin-orbit coupling (SOC) term $\mathscr{H}_\textrm{SOC}$, and the effective Coulomb interaction term $\mathscr{H}_\textrm{C}$. Surrounding atoms of the \textit{RE} atom affect not only $\mathscr{H}_\textrm{CF}$ but also $\mathscr{H}_\textrm{C}$ through the size of interaction, and $\mathscr{H}_\textrm{SOC}$ as well via the shape of $4f$ orbitals. $\mathscr{H}_\textrm{SOC}$ and $\mathscr{H}_\textrm{C}$ are given as
\begin{align}
    \mathscr{H}_\textrm{SOC}&=\xi\sum_{\substack{mm'\\\sigma\sigma'}}
    (l^{mm'}_xs^{\sigma\sigma'}_x+l^{mm'}_ys^{\sigma\sigma'}_y+l^{mm'}_zs^{\sigma\sigma'}_z)
    \hat{c}^{\dagger}_{m\sigma}\hat{c}_{m'\sigma'},\\
    \mathscr{H}_\textrm{C}&=\frac{1}{2}\sum_{\substack{m_1,m_2,\\m_3,m_4}}\sum_{\sigma\sigma'}g^{\sigma\sigma'}_{m_1 m_2 m_3 m_4}\hat{c}^{\dagger}_{m_1 \sigma}\hat{c}^{\dagger}_{m_2 \sigma'}\hat{c}_{m_4 \sigma'}\hat{c}_{m_3\sigma}.
\end{align}
Here indices $m, m', m_i (i=1,2,3,4)$ are for the magnetic quantum number, $\sigma$ and $\sigma'$ for spins, and $\hat{c}_{m\sigma}$ is the electron-annihilation operator. $\mathscr{H}_\textrm{SOC}$ is made of the strength of SOC $\xi$ and the angular-momentum and spin matrices $l_x^{mm'}$, $s_x^{\sigma\sigma'},\cdots$. We assume the effective Coulomb interactions $g^{\sigma\sigma'}_{m_1 m_2 m_3 m_4}$ are given as
\begin{align}
    g^{\sigma\sigma'}_{m_1 m_2 m_3 m_4}=&(-1)^{m_1-m_3}\delta_{m_1+m_2,m_3+m_4}\nonumber\\
        &\times\sum^{l}_{p=0}F^{2p}_{\sigma\sigma'}c^{2p}(m_1,m_3)c^{2p}(m_2,m_4),
\end{align}
where we have the Gaunt coefficients $c^{p}(m,m')$ \cite{doi:10.1098/rsta.1929.0004} and the Slater-Condon parameters $F^{2p}_{\sigma\sigma'}$ for $4f$ orbitals\cite{PhysRev.34.1293,PhysRev.37.1025}. We use the scaled-Slater-Condon parameters $F_0, \varDelta F_0$ and $F_2$ to represent $F^{2p}_{\sigma\sigma'}$ as  
\begin{align}
    \left\{
        \begin{aligned}
            F^0_{\sigma\sigma'}=&F_0+ \delta_{\sigma\sigma'}\varDelta F_0\\
            F^2_{\sigma\sigma'}=&225F_2\\
            F^4_{\sigma\sigma'}=&1089\times 0.138F_2\\
            F^6_{\sigma\sigma'}=&7361.64\times 0.151F_2
        \end{aligned}
    \right. 
\label{eq:scratio}
\end{align}
in the manner of Ref.~\cite{Trefftz1951} for analyzing the \textit{RE}$^{3+}$ elements. Here we fix the ratios $F^4_{\sigma\sigma'}/F^2_{\sigma\sigma'}$ and $F^6_{\sigma\sigma'}/F^2_{\sigma\sigma'}$ as in Ref.\cite{doi:10.1098/rspa.1955.0037} respecting the case of Hydrogen. The spin-dependent term $\delta_{\sigma\sigma'}\varDelta F_0$ is introduced to make a compromise in our fitting procedure to the QSGW results with keeping the spin-space symmetry of the model Hamiltonian.

We assume a general crystal field that corresponds to the symmetry of each material. Using Steven's operators $(O^0_4)_{mm'},(O^4_4)_{mm'},$ and so on\cite{Stevens_1952}, $\mathscr{H}_\textrm{CF}$ is given as 
\begin{align}
    \mathscr{H}_\textrm{CF}&=\sum_{\substack{mm'\\\sigma\sigma'}} (h_\textrm{CF})_{mm'}\hat{c}^{\dagger}_{m,\sigma}c_{m',\sigma'}\\
    h_\textrm{CF}&=
    \begin{cases}
        B^0_4(O^0_4+5O^4_4)+B^0_6(O^0_6-21O^6_6)&\textrm{(cubic)}\\
        B^0_2O^0_2+B^0_4O^0_4+B^0_6O^0_6+B^6_6O^6_6 &\textrm{(hexagonal)}
    \end{cases},
\end{align}
where $B^0_4, B^0_6, B^0_2, B^6_6$ are the parameters to specify crystal fields. Thus, the model Hamiltonian $\mathscr{H}$ of Eq.~(\ref{eq:hamiltonian}) is specified by parameters $\xi, F_0, \varDelta F_0, F_2$, and $B^m_l$.

To apply our core idea shown in the introduction, we should apply QSGW to the model Hamiltonian. Here we neglect the correlation part since we expect little screening effects of $4f$ orbitals by themselves. That is, we apply the Hartree-Fock approximation to $\mathscr{H}$ instead, resulting in the Hartree-Fock model Hamiltonian (HFMH) $\mathscr{H}_\textrm{HF}$ as
\begin{align}
    \mathscr{H}_\textrm{HF}=&\mathscr{H}_\textrm{0}+\mathscr{H}_\textrm{SOC}+\mathscr{H}_\textrm{CF}+\mathscr{H}^\textrm{HF}_\textrm{C}\nonumber\\
    \mathscr{H}^\textrm{HF}_\textrm{C}=&\sum_{m_1,m_3}\sum_{\sigma}\left[\sum_{m_2,m_4}(g_{m_1 m_2 m_3 m_4}^{\sigma\sigma}
-g_{m_1 m_2 m_4 m_3}^{\sigma\sigma})\braket{c^{\dagger}_{m_2 \sigma}c_{m_4 \sigma}}\right.\nonumber\\
    &\left.\hspace{5em}+g_{m_1 m_2 m_3 m_4}^{\sigma\overline{\sigma}}\braket{c^{\dagger}_{m_2 \overline{\sigma}}c_{m_4 \overline{\sigma}}}\right]\hat{c}^{\dagger}_{m_1 \sigma}\hat{c}_{m_3 \sigma},
\label{eq:hmf}
\end{align}
where $\mathscr{H}^\textrm{HF}_\textrm{C}$ is the mean-field approximation to $\mathscr{H}_\textrm{C}$. $\overline{\sigma}$ denotes the opposite spin to $\sigma$, and $\braket{\dots}$ means the expectation values for the ground state. 
Based on our core idea, we compare $\mathscr{H}_\textrm{HF}$ with the $4f$ part of the one-body Hamiltonian 
$\mathcal{H}^\textrm{QSGW}_{4f}$ given in QSGW, so as to determine parameters in $\mathscr{H}$.

We use the ecalj package\cite{ecalj} to perform QSGW calculations. 
In practice, we use $20\%$ LDA mixing (QSGW80) so as to reduce too-large exchange effects. 
We can take QSGW80 as a quick remedy to include the effects of the vertex correction, 
which enhances the screening effect by $\sim20$\%\cite{sakakibara_finite_2020,sakakibara_model-mapped_2017}.
In fact, QSGW80 can reproduce the experimental band gap very well\cite{1347-4065-55-5-051201,PhysRevLett.96.086405,1347-4065-55-5-051201}. 
We obtain $\mathcal{H}^\textrm{QSGW}_{4f}$ based on localized Wannier functions (it was implemented in ecalj.) as $\langle w_m|\mathscr{H}^{\rm QSGW}_{4f}| w_{m'}\rangle$.
Here $w_m$ and $w_{m'}$ are the Wannier functions generated 
by the one-shot projection of $4f$-type atomic-like seed orbitals \cite{PhysRevB.56.12847,ecalj}.

In the QSGW calculations, we add the SOC term $\mathscr{H}^{\rm QSGW}_\textrm{SOC}$ in the $L_z S_z$-only approximation. We obtain $\xi$ by an average
\begin{align}
\xi=\sqrt{\left(\sum_n |\Bra{w_n}\mathscr{H}^{\rm QSGW}_\textrm{SOC}\Ket{w_n}|^2\right)/14}.
\end{align}
Other parameters $F_0, \varDelta F_0, F_2$, and $B^m_l$ of $\mathscr{H}$ are determined to minimize the difference of eigenvalues between $\mathcal{H}^\textrm{QSGW}_{4f}$ and $\mathscr{H}_\textrm{HF}$. Then, we can obtain eigenvalues of $\mathscr{H}$ by the exact diagonalization \cite{mycodes}.

In a summary, we mainly added two approximations to our core idea. One is $\mathscr{H}_\textrm{HF}$ instead of applying QSGW to the model Hamiltonian. The other is the QSGW80 instead of QSGW. In addition, we fix the ratios $F^4_{\sigma\sigma'}/F^2_{\sigma\sigma'}$ and $F^6_{\sigma\sigma'}/F^2_{\sigma\sigma'}$. We utilize the fixed ratio to enhance numerical stability and simple interpretation. 
In principle, our core idea is rather general for extracting an essential degree of freedom from the first-principles calculations. One of the advantages of our core idea is that we do not calculate effective interaction directly. The calculation of effective interaction is somehow complicated as discussed in Refs.~\onlinecite{sakakibara_finite_2020,sakakibara_model-mapped_2017}, especially when we like to include vertex corrections.

\section{results}
Prior to discussion of the \textit{RE} in compounds, we have examined free $RE^{+3}$ in QSGW to confirm the performance of our method. For the calculation of the free $RE^{+3}$, we use a fcc supercell placing $RE^{+3}$ at their centers, where $RE^{+3}$ are separated by $7.07$~\AA.\ We assume a homogeneous background charge to keep charge neutrality. 

\begin{figure*}[htbp]
    \centering
    \includegraphics[width=0.9\hsize,clip]{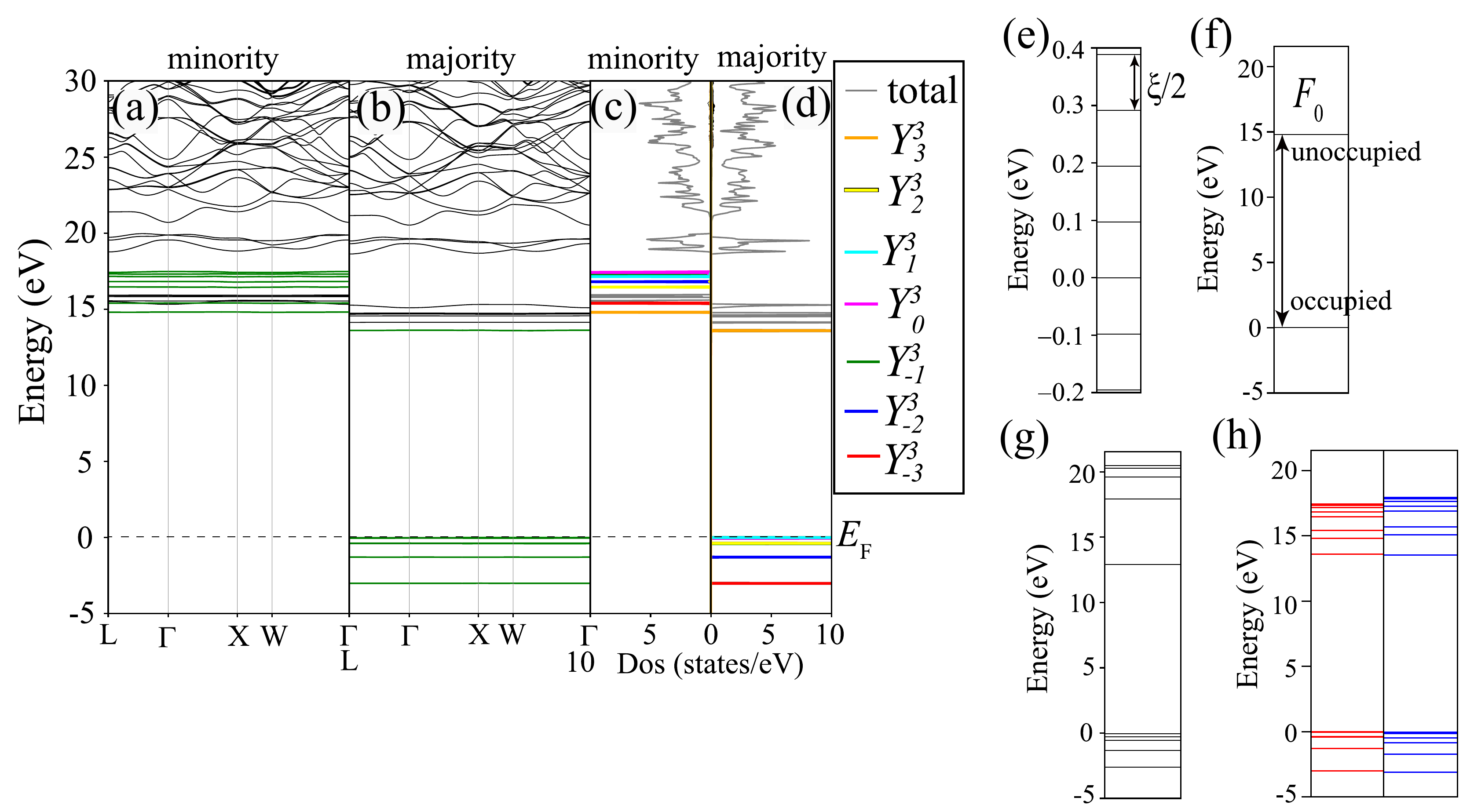}
\caption{(a),(b) Band structure of QSGW (black) and $\mathcal{H}^\textrm{QSGW}_{4f}$ (green) for free Eu$^{3+}$ in a fcc supercell. (c),(d) DOS and PDOS corresponding to the QSGW band structure. (e)-(h) are for analyzing contributions of parameters
$\xi, F_0, \varDelta F_0$, and $F_2$ to $\mathscr{H}_\textrm{HF}$. We show eigenvalues; (e) only with $\xi$;(f) only with $F_0$; and (g) only with $F_0$ and $F_2$. (h) Comparison of eigenvalues of $\mathcal{H}^\textrm{QSGW}_{4f}$(red) and $\mathscr{H}_\textrm{HF}$  (blue). Spin-dependence is not resolved in (e)-(h).}\label{ionband}
\end{figure*}

In Figs.~\ref{ionband}~(a)~and~\ref{ionband}~(b), we show the calculated band structure of Eu$^{3+}$ in QSGW. We superpose that of $\mathcal{H}^\textrm{QSGW}_{4f}$ with green flat lines (seven lines per spin).
PDOS of $4f$ orbitals is shown in Figs.~\ref{ionband}~(c) and (d). Up to $\sim 18$~eV, the calculated bands corresponding to $4f$, $5d$, and $6s$ are almost flat, indicating that our supercell is large enough. On the other hand, $6p$ bands around $\sim 20$~eV have band width $\sim 1$~eV because of the finite size of our supercell. The eigenvalue at $\Gamma$ point around $\sim 21.5$~eV is identified to be the vacuum level, the bottom of the scattering states. The band gap (LUMO-HOMO gap) is 13.7~eV, much larger than the LDA value of $4.85$~eV. The majority bands of $4f$ orbitals except for $m=3$ are occupied, that is, the ground state is $J_z=L_z+S_z=-3+3=0$, corresponding to $^{7}\textrm{F}_{0}$. Thus, our results are consistent with Hund's rule. Generally speaking, a good mean-field approach should satisfy Hund's rule since the ground states of atoms should be essentially described by the electronic configuration of a single Slater determinant.

We determine $F_0$, $\varDelta F_0$, and $F_2$ in $\mathscr{H}_\textrm{HF}$ so as to reproduce eigenvalues of $\mathcal{H}^\textrm{QSGW}_{4f}$. 
We see the eigenvalues match well as shown in Fig.~\ref{ionband}~(h). Figures.~\ref{ionband}~(e)-(h) show our analysis of how the parameters affect the eigenvalues of $\mathscr{H}_\textrm{HF}$. Figure~\ref{ionband}~(e) shows $\xi$ splits into seven states with degeneracy between $m\sigma$ and $-m\overline{\sigma}$. Figure~(f) shows that $F_0$ makes the difference between occupied and unoccupied states. Figure~(g) shows that $F_0$ with $F_2$ is still not enough to reproduce $\mathcal{H}^\textrm{QSGW}_{4f}$. The parameter $\varDelta F_0$ is for describing larger effective interaction between occupied orbitals than that between occupied and unoccupied orbitals. This is reasonable because occupied $4f$ orbitals are localized more than unoccupied orbitals.

In Table~\ref{tabxifn}, we summarize obtained $\xi$, $F_0$, $\varDelta F_0$, and $F_2$ for \textit{RE}$^{+3}$ in our method. The QSGW results are in the Supplemental Material\cite{supplemental}. We skipped Ce since QSGW did not give localized $4f$ eigenfunctions. Table~\ref{tabxifn} shows that $\xi$ and $F_2$ in our method give good agreements with those of empirical studies: their differences are only $\sim$10\%. This is a justification for our method.
$F_0$, corresponding to $U$ of LDA+$U$, is changing systematically along the atomic number. Except for Gd, $F_0$ is the largest in the middle of the first half and the latter half of the $4f$ series, namely at Sm when filling majority electrons, and at Er when minority. The latter half of the species shows a little larger values than the first half. $F_2$ systematically becomes larger along the atomic number. For example, the $F_2$ values of Tm and Pr are 0.080 eV and 0.043 eV, and the difference is 0.037 eV in our method, while the empirical study shows these $F_2$ values are 0.056 eV and 0.038 eV, and the difference is as 0.018 eV.

\begin{figure}[htbp]
    \centering
    \includegraphics[width=8cm,clip]{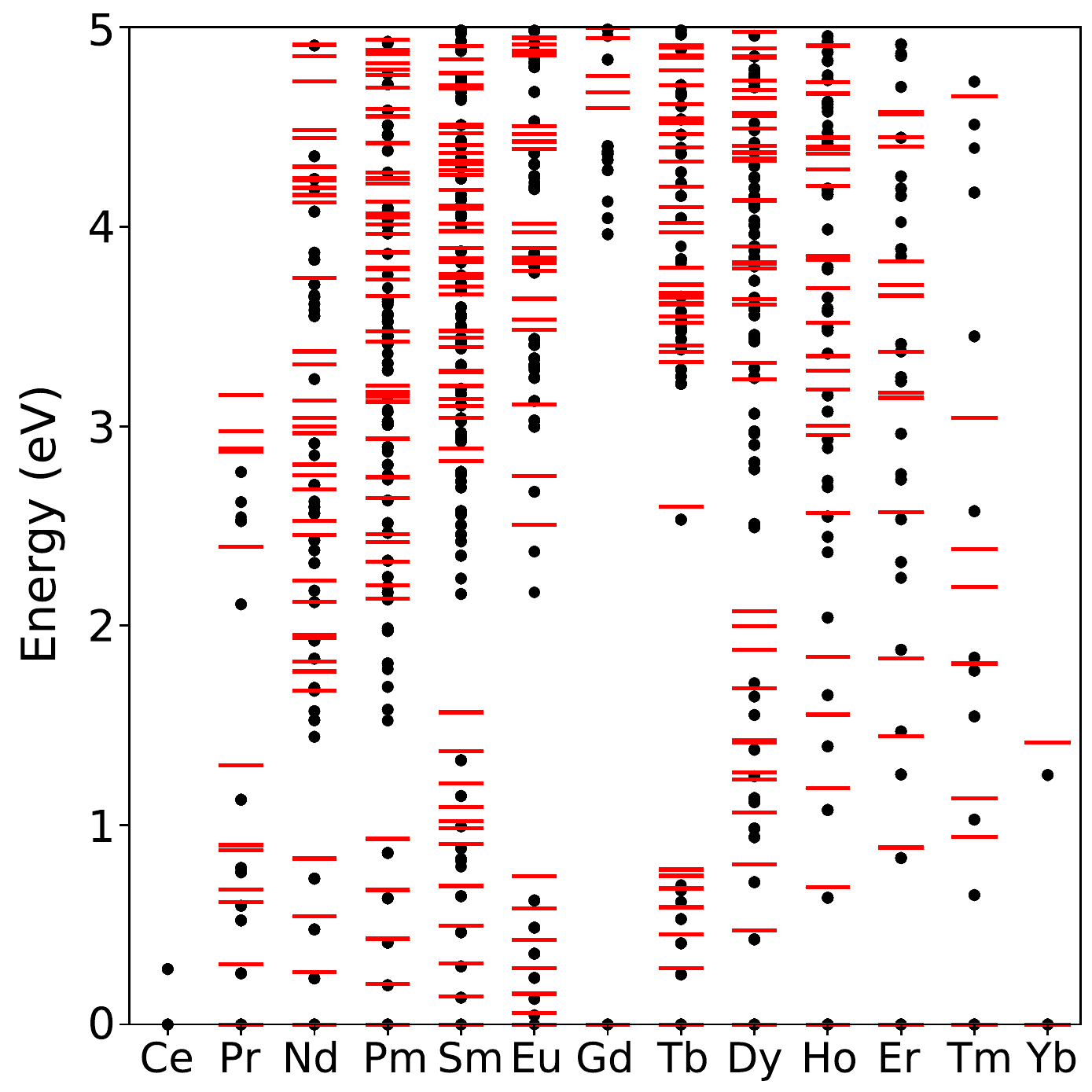}
    \caption{Eigenvalues of \textit{RE} ions. Dieke diagram is shown by the black points. Our results are superposed by red lines. Both are calculated from parameters in Table \ref{tab:triv} }
    \label{ddia}
\end{figure}

In Fig.~\ref{ddia}, we compare the excitation energies of free \textit{RE}$^{3+}$ by the parameters of our method and by those of empirical studies in Table~\ref{tabxifn}. The latter exactly reproduced the original Dieke diagram\cite{Dieke:63}. In free \textit{RE}$^{3+}$, the excitation energies only depend on $\xi$ and $F_2$. Although the lowest excitation energies show good agreements (except for Tm with a little large error of $\sim$30\%), we see disagreements overall. This is because the excitation energies are somehow sensitive to the small differences of $\xi$ and $F_2$ in Table~\ref{tabxifn}.
\begin{table}[tbhp]
    \centering
    \caption{Parameters in eV determined in our method for free trivalent ions. The empirical studies in the right column correspond to the empirical values of Dieke's review paper\cite{Dieke:63}. These values are derived from many experimental and theoretical data\cite{doi:10.1139/cjr35a-001, doi:10.1139/cjr36a-013,doi:10.1098/rspa.1959.0097,doi:10.1063/1.1730954,doi:10.1063/1.1730267,doi:10.1063/1.1732054,doi:10.1063/1.1732059,doi:10.1063/1.1732879,doi:10.1063/1.1732054,doi:10.1063/1.1733947,doi:10.1063/1.1732407}.}\label{tabxifn}
    \begin{tabular}{|c||rccc|ll|}
        \hline
        &\multicolumn{4}{c|}{Our method} &\multicolumn{2}{c|}{Empirical study}\\
        &$F_0$ & $\varDelta F_0$ & $\xi$ & $F_2$ & $\xi$ & $F_2$\\
        \hline
        Pr& 16.107 &  0.025 & 0.106 & 0.043 & 0.091\cite{doi:10.1063/1.1732054} & 0.038\cite{doi:10.1063/1.1732054}\\
        Nd& 16.573 &  0.108 & 0.123 & 0.049 & 0.109 & 0.042\\
        Pm& 16.672 &  0.282 & 0.143 & 0.062 & 0.133\cite{doi:10.1063/1.1732059} & 0.043\cite{doi:10.1063/1.1732059}\\
        Sm& 17.479 &  0.352 & 0.162 & 0.059 & 0.149 & 0.046\cite{doi:10.1063/1.1732879}\\
        Eu& 14.803 &  0.481 & 0.195 & 0.058 & 0.164\cite{doi:10.1063/1.1733947} & 0.050\cite{doi:10.1063/1.1733947}\\
        Gd& 17.317 &  0.000 & 0.208 & 0.058 & 0.196 & 0.050\\
        Tb& 15.740 & -0.274 & 0.236 & 0.055 & 0.211\cite{doi:10.1063/1.1733947} & 0.054\cite{doi:10.1063/1.1733947}\\
        Dy& 16.867 &  0.000 & 0.265 & 0.068 & 0.236 & 0.052\cite{doi:10.1063/1.1732879}\\
        Ho& 16.817 &  0.000 & 0.296 & 0.071 & 0.268\cite{doi:10.1063/1.1732059} & 0.056\cite{doi:10.1063/1.1732059}\\
        Er& 16.905 &  0.000 & 0.329 & 0.078 & 0.303\cite{doi:10.1063/1.1732407} & 0.053\cite{doi:10.1063/1.1732407}\\
        Tm& 16.228 &  0.192 & 0.365 & 0.080 & 0.329 & 0.056\\
        Yb& 10.610 &  0.000 & 0.404 & 0.073 & 0.357 & ~~~~-  \\
        \hline
    \end{tabular}
    \label{tab:triv}
\end{table}

Let us show our main results of Eu compounds, EuCl$_3$, EuN, and Eu-doped GaN. For Eu-doped GaN in the wurtzite structure, we use a $2\times 2\times 2$ supercell (32 atoms per cell) substituting one Ga with Eu. We perform QSGW calculations for the crystal structure optimized by Quantum Espresso\cite{0953-8984-21-39-395502,optparas}. We show the QSGW results for EuCl$_3$ and EuN in Supplemental Material\cite{supplemental}.
In Table~\ref{tabxifneeu}, we show $\xi$, $F_0$, $\varDelta F_0$, $F_2$, and $B^m_l$. We see $F_0$ and $\varDelta F_0$ are strongly reduced from those of the free Eu ion. However, $F_2$ is not so different from that of free Eu ion. This finding suggests that $F_2$ given by Dieke is reasonable probably even in other solids. However, we see some complicated behavior of $F_2$ in detail: 
$F_2$=0.058~eV for the free Eu ion is reduced to be $F_2$=0.053 eV for EuN while a little enhanced to be $F_2$=0.075~eV for EuCl$_3$.
On the other hand, $\xi$=0.195~eV for free Eu ion is very different from $\xi$=0.093~eV for Eu-doped GaN. Considering the expression of SOC, this is due to the delocalization of $4f$ orbitals in Eu-doped GaN, suggesting hybridization with other surrounding orbitals. The parameters for CF should be material dependent. In particular, we see that $B^0_2$ for the $c$-axis anisotropy is rather large (=2.217~eV) for EuCl$_3$, while $B^0_2$=0.346~eV for Eu-doped GaN.

\begin{figure*}[hbtp]
    \centering
    \includegraphics[width=0.8\hsize,clip]{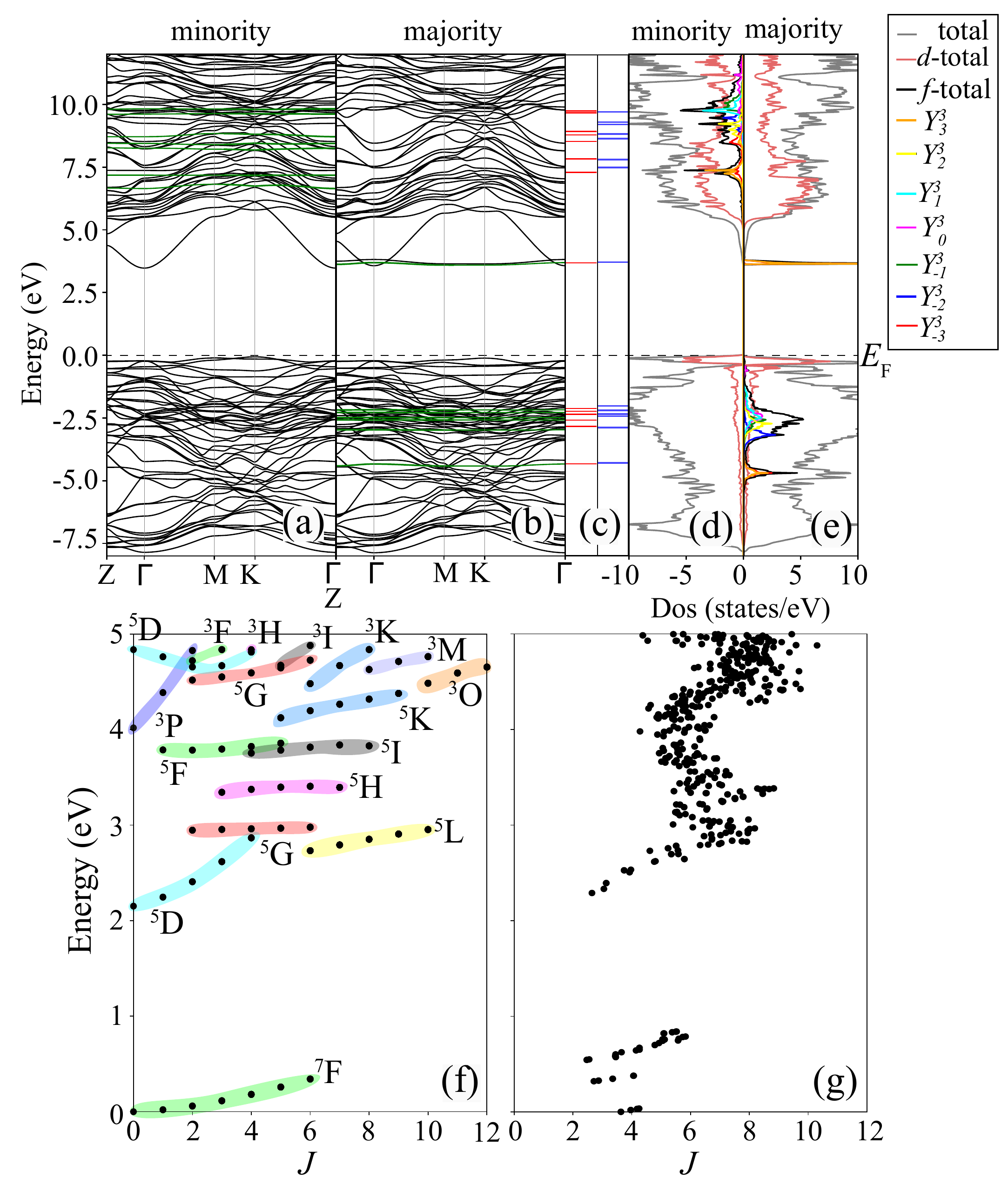}
    \caption{(a),(b) The QSGW band structure of majority (minority) spin for Eu-doped GaN (black). We superpose band structure of  $\mathcal{H}^\textrm{QSGW}_{4f}$ (green). 
    (c) Comparison of eigenvalues of $\mathcal{H}^\textrm{QSGW}_{4f}$ (red) and those of $\mathscr{H}_\textrm{HF}$ (blue). (d),(e) DOS and PDOS corresponding to the QSGW band structure. (f) The plot of eigenvalues of $\mathscr{H}$ for Eu-doped GaN classified by $J$ neglecting $\mathscr{H}_\textrm{CF}$. (g) The plot of eigenvalues of $\mathscr{H}$ for Eu-doped GaN classified by $J$.}
    \label{figEuCompound}
\end{figure*}

\begin{table*}[htbp]
    \centering
    \caption{Parameters in eV determined by our method for Eu$^{3+}$ in Eu compounds. See text for their definitions. For the calculations, we use crystal structures in references.}\label{tabxifneeu}
    \begin{tabular}{c|crcccccc}
        \hline
        material                            & $\xi$ & $F_0$ &$\varDelta F_0$& $F_2$ & $B^0_4$ & $B^0_6$ & $B^0_2$ & $B^6_6$\\
        \hline\hline
        Free Eu ion (Table \ref{tab:triv})  & 0.195 & 14.803 & 0.481 & 0.058 &   -   &    -   &   -   &   -   \\
        EuCl$_3$\cite{osti_1275324}         & 0.203 &  8.773 & 0.340 & 0.075 & 0.017 &  0.004 & 2.217 & -0.080\\
        EuN\cite{PhysRevB.75.045114}        & 0.156 &  5.565 & 0.150 & 0.053 & 0.000 & -0.006 &   -   &    -  \\
        Eu-doped GaN                        & 0.093 &  6.703 &-0.057 & 0.046 & 0.021 & -0.002 & 0.346 & -0.063\\
        \hline
    \end{tabular}
\end{table*}
Let's focus on Eu-doped GaN. As shown in Figs.~\ref{figEuCompound}~(a) and (b), QSGW80 gives the band gap 3.78 (3.69) eV for the majority (minority) spin, corresponding to the experimental value of GaN 3.4~eV. Just above the bottom of the conduction band, we have an unoccupied band of $m=\pm3$ for the majority spin. In contrast to the unoccupied band, the occupied $4f$ orbitals are hybridized well with valence bands of GaN. 
Such hybridization is also seen in the calculations of the HSE functional \cite{PhysRevMaterials.6.044601}.
In Fig.~\ref{figEuCompound}~(c), we compare the eigenvalues of $\mathscr{H}_\textrm{HF}$ (blue lines) and those of $\mathcal{H}^\textrm{QSGW}_{4f}$ (red lines). A little matching error indicates room for improving our method, while the error may not change our conclusions.

We show the eigenvalues of $\mathscr{H}$ in Fig.~\ref{figEuCompound}~(g) obtained by the exact diagonalization. Here we classify the eigenvalues by the expectation values of total angular momentum $J$. In Fig.~\ref{figEuCompound}~(f), we show eigenvalues when we neglect $\mathscr{H}_\textrm{CF}$ in $\mathscr{H}$. Without $\mathscr{H}_\textrm{CF}$, each eigenstate is represented by Russell-Saunders states. In Fig~\ref{figEuCompound}~(f) without CF, there is a large gap between $^{7}$F and $^{5}$D. The size of this gap is $\sim 2$~eV. This corresponds to the observed red emission which is experimentally identified to be the transition from $^5$D$_0$ to $^7$F$_2$\cite{doi:10.1063/1.3478011}. Comparison of Figs.~\ref{figEuCompound}~(f) and (g) shows that $\mathscr{H}_\textrm{CF}$ is large enough to mix low-$J$ and high-$J$ eigenstates, resulting in a middle size of $J$. That is, $\mathscr{H}_\textrm{CF}$ causes a sizable breaking of the Russell-Saunders coupling. However,  $\mathscr{H}_\textrm{CF}$ is not large enough to alter the overall structure. That is, we see remnants of $^5$D and $^7$F in Fig.~\ref{figEuCompound}~(g) while the excitation gap between them is preserved. This justifies the use of the Dieke diagram for the analysis of PLR.

In summary, we presented a new method to determine the model Hamiltonian from the QSGW calculations. By the exact diagonalization of the model Hamiltonian, we can justify the applicability of the Dieke diagram to \textit{RE} in solids and its limitations. Along the line of our method, it is possible to include the hybridization of $4f$ electrons with others. In fact, our QSGW calculation shows we have large hybridization of occupied $4f$ orbitals with valence bands. We now have an extension of our method applied to multiplets of 3$d$ orbitals working well \cite{saito}.

We thank Y.~Fujiwara, S.~Ichikawa, J.~Tatebayashi, D.~Timmerman, M.~Ashida, H.~Ishihara, H.~Ikeda, H.~Kusunose, T.~Oda, H.~Usui, and H.~Okumura for valuable discussions. This study is partly supported by JSPS KAKENHI (Grants No. 18H05212, No. 20K05303, No. 22K04909).

\bibliography{GaN}

\begin{thebibliography}{55}
\bibitem [{opt()}]{optparas}%
  \BibitemOpen
  \href@noop {} {}\bibinfo {note} {We use the wave functions expanded by plane
  waves up to cut-off energy of 60.0 Ry and $8^3$ k-point mesh for
  optimization.}\BibitemShut {Stop}%
\end{thebibliography}
\end{document}


\title{First-principles method justifying the Dieke diagram: Supplemental material}
\author{Katsuhiro~Suzuki} 
\affiliation{Division of Materials and Manufacturing Science, Graduate School of Engineering, Osaka University, Suita, Osaka 565-0871, Japan}
\author{Takao~Kotani}
\affiliation{Advanced Mechanical and Electronic System Research Center (AMES), Faculty of Engineering, Tottori University, Tottori 680-0945, Japan}
\affiliation{CSRN-Osaka, Osaka University, Toyonaka, Osaka 560-8531, Japan}
\author{Kazunori~Sato}
\affiliation{Division of Materials and Manufacturing Science, Graduate School of Engineering, Osaka University, Suita, Osaka 565-0871, Japan}
\affiliation{CSRN-Osaka, Osaka University, Toyonaka, Osaka 560-8531, Japan}
\affiliation{Spintronics Research Network Division, OTRI, Osaka University, Toyonaka, Osaka 560-8531, Japan}
\date{\today}
\maketitle
\begin{figure*}[hbtp]
    \centering
    \includegraphics[width=16cm,clip]{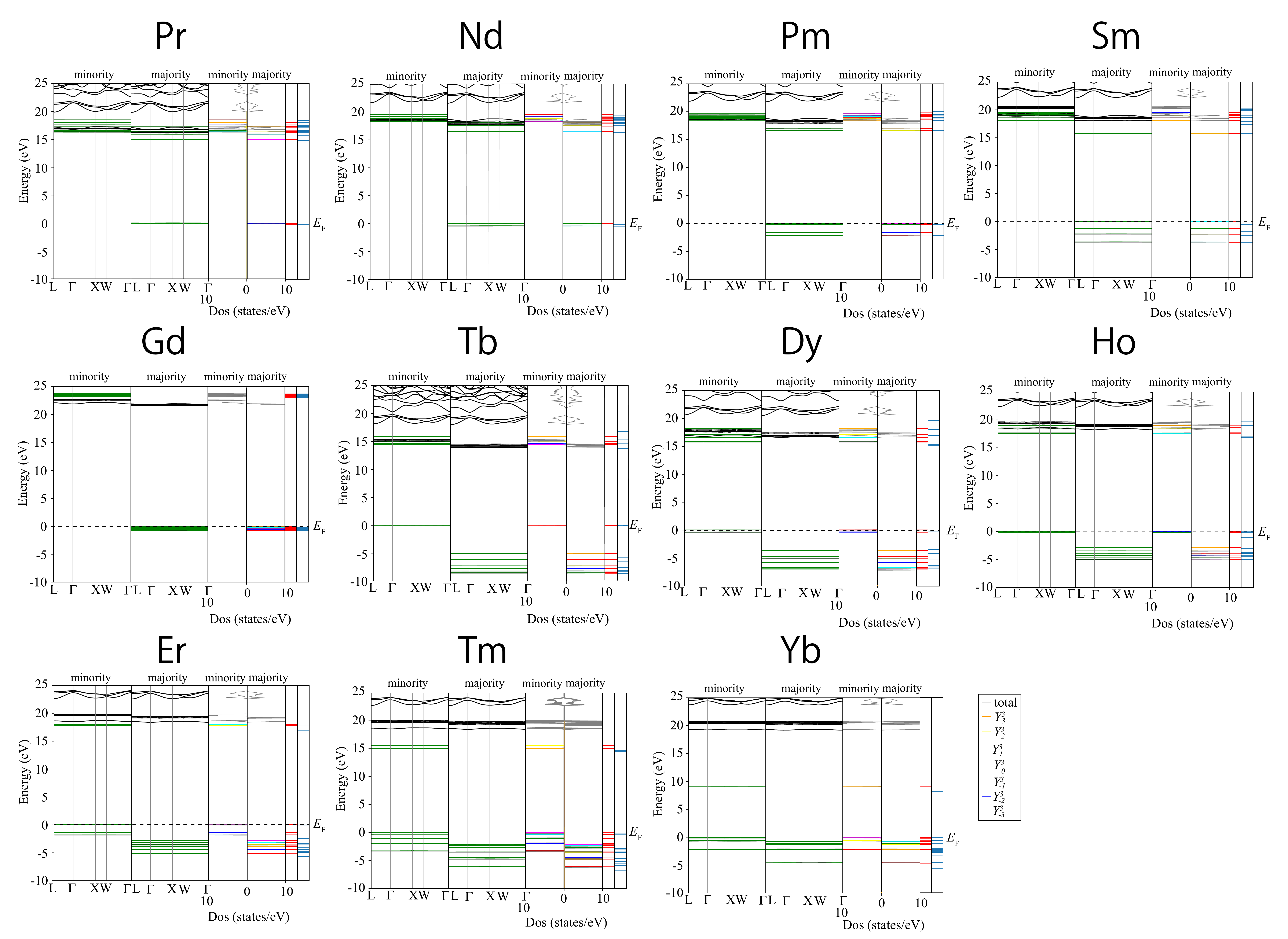}
    \caption{(left) Band structure of QSGW (black), 14-orbital model from MLWFs (green), (middle) DOS and PDOS, and (right) comparison of eigenvalues of 14-orbital model (red) and HFMH (blue) of each \textit{RE} ions.}
    \label{figions}
\end{figure*}

In Supplemental material, we show the results which cannot be included in text. First, we show the band structure and density of states of free \textit{RE} ions in Fig.~\ref{figions}. We take $8\times 8\times 8$ $k$-mesh for scf calculation and $2\times 2\times 2$ $q$-mesh for self-energy.

\begin{figure}[hbtp]
    \centering
    \includegraphics[width=8cm,clip]{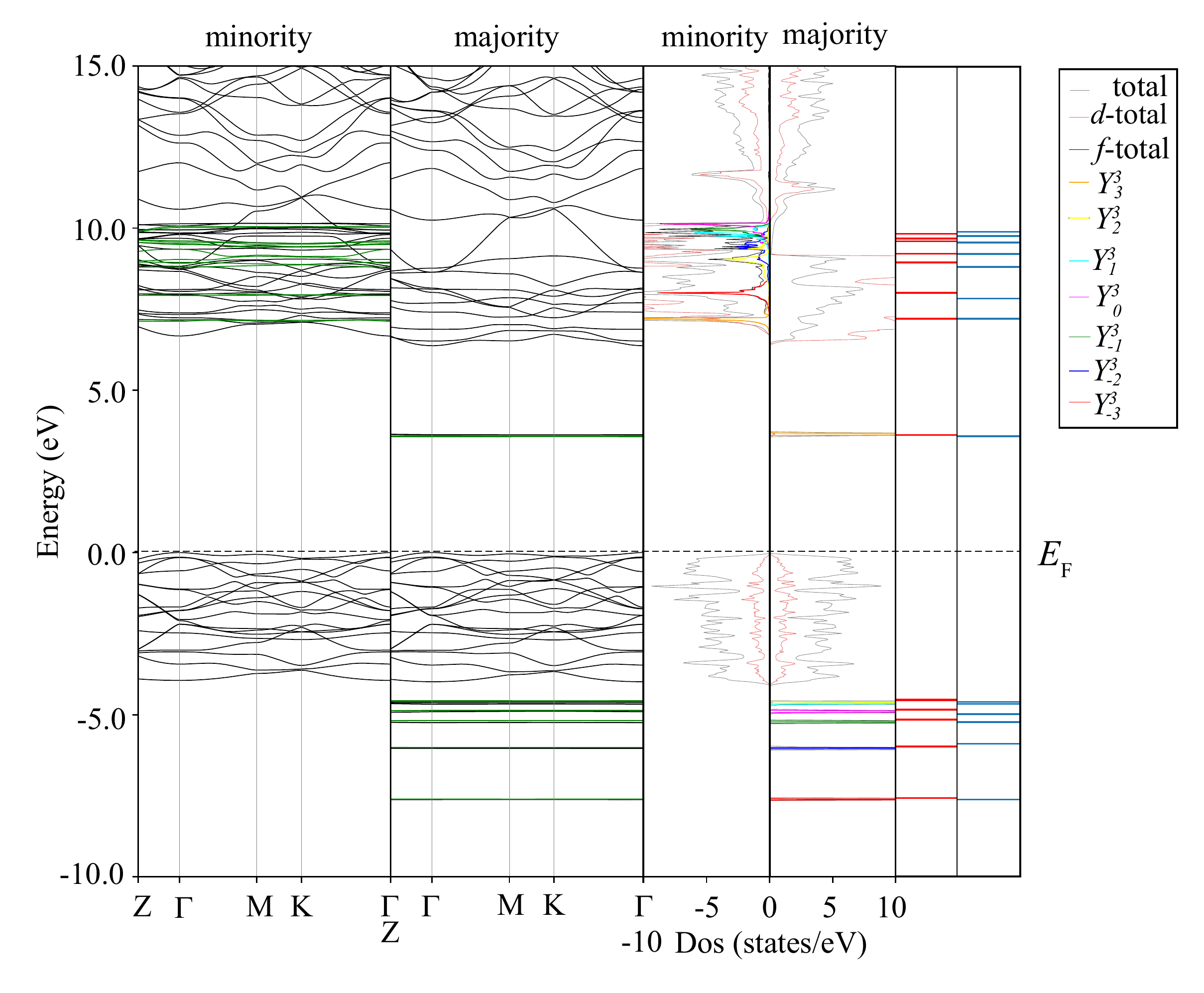}
    \caption{(left) Band structure of QSGW (black) and 14-orbital model from MLWFs (green), (middle) DOS and PDOS, and (right) comparison of eigenvalues of 14-orbital model (red) and HFMH (blue) of EuCl$_3$.}
    \label{figEuCl3}
\end{figure}

Next, we show the band structure and Comparison of eigenvalues of 14-orbital model and HFMH. Fig.~\ref{figEuCl3} shows that of EuCl$_3$. We apply crystal structure as Ref.~\onlinecite{osti_1275324}. We take $8\times 8\times 8$ $k$-mesh for scf calculation and $2\times 2\times 2$ $q$-mesh for self-energy. QSGW80 gives band gap $6.4\sim 6.7$eV. 

\begin{figure}[hbtp]
    \centering
    \includegraphics[width=8cm,clip]{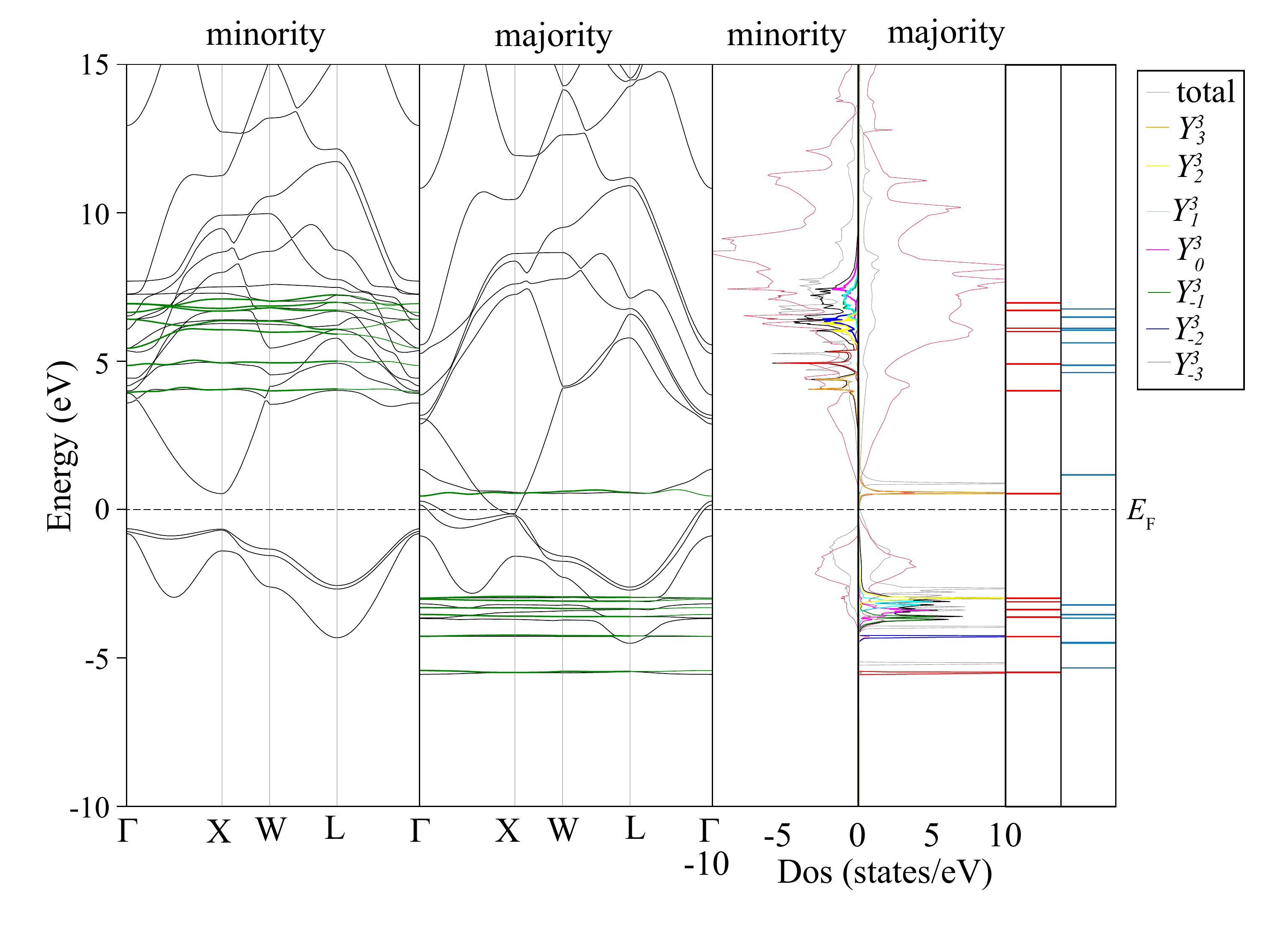}
    \caption{(left) Band structure of QSGW (black) and 14-orbital model from MLWFs (green), (middle) DOS and PDOS, and (right) comparison of eigenvalues of 14-orbital model (red) and HFMH (blue) of EuN.}
    \label{figEuN}
\end{figure}

In Fig.~\ref{figEuN}, Band structure of EuN is shown. We take $6\times 6\times 6$ $k$-mesh for scf calculation and $6\times 6\times 6$ $q$-mesh for self-energy. we use crystal structure as Ref.~\onlinecite{PhysRevB.75.045114}. QSGW80 gives band gap $0\sim 1.2$eV. 
\bibliographystyle{apsrev4-2}
\bibliography{GaN}